\documentclass[conference]{IEEEtran}
\IEEEoverridecommandlockouts
\usepackage{cite}
\usepackage{amsmath,amssymb,amsfonts}
\usepackage{algorithmic}
\usepackage{graphicx}
\usepackage{textcomp}
\usepackage{xcolor}

\usepackage{caption}
\usepackage{subcaption}
\usepackage{multirow}
\usepackage{stfloats}

\def\BibTeX{{\rm B\kern-.05em{\sc i\kern-.025em b}\kern-.08em
    T\kern-.1667em\lower.7ex\hbox{E}\kern-.125emX}}
\begin{document}

\title{Shadow Augmentation for Handwashing Action Recognition: from Synthetic to Real Datasets\\
}

\author{\IEEEauthorblockN{Shengtai Ju, Amy R. Reibman}
\IEEEauthorblockA{\textit{Elmore Family School of Electrical and Computer Engineering} \\
\textit{Purdue University}\\
West Lafayette, USA \\
ju10@purdue.edu, reibman@purdue.edu}
}

\maketitle

\begin{abstract}
Video analytics systems designed for deployment in outdoor conditions can be vulnerable to many environmental changes, particularly changes in shadow. Existing works have shown that shadow and its introduced distribution shift can cause system performance to degrade sharply. In this paper, we explore mitigation strategies to shadow-induced breakdown points of an action recognition system, using the specific application of handwashing action recognition for improving food safety. Using synthetic data, we explore the optimal shadow attributes to be included when training an action recognition system in order to improve performance under different shadow conditions. Experimental results indicate that heavier and larger shadow is more effective at mitigating the breakdown points. Building upon this observation, we propose a shadow augmentation method to be applied to real-world data. Results demonstrate the effectiveness of the shadow augmentation method for model training and consistency of its effectiveness across different neural network architectures and datasets. 

\end{abstract}


\section{Introduction}
\label{sec:intro}
For outdoor camera-based systems and applications, the presence of shadow is natural and inevitable. The properties of natural shadow, such as the size and intensity, can vary depending on time of the day and placement of the camera system, which create challenges for designing a robust image/video system. However, system robustness against different shadow conditions is crucial for successful camera-system applications in the outdoors. Despite the significance of shadow, quantitative studies of how shadow impacts a recognition system and how to deal with performance degradation caused by shadow are still lacking.
In our previous work \cite{ju2023robust}, we demonstrated that changes in environmental conditions, including shadow conditions, caused system robustness to drop significantly, where system robustness was measured as the classification accuracy on unseen datasets with distribution shift. Distribution shift, which occurs when training and testing data are sampled from different distributions, has been shown to impact system performance for the task of image classification \cite{recht2019imagenet, ood_bench}. 
Therefore, in this paper, we explore methods to alleviate system performance degradation caused by shadow, and we improve system robustness with shadow augmentation in the context of action recognition. More specifically, we have chosen the application of handwashing action recognition for better hand hygiene and improved food safety \cite{zhong2019hand, zhong2020multi, zhong2021designing, ju2024mipr}.

\begin{figure}[htb!]
\centering
\includegraphics[width=\linewidth]{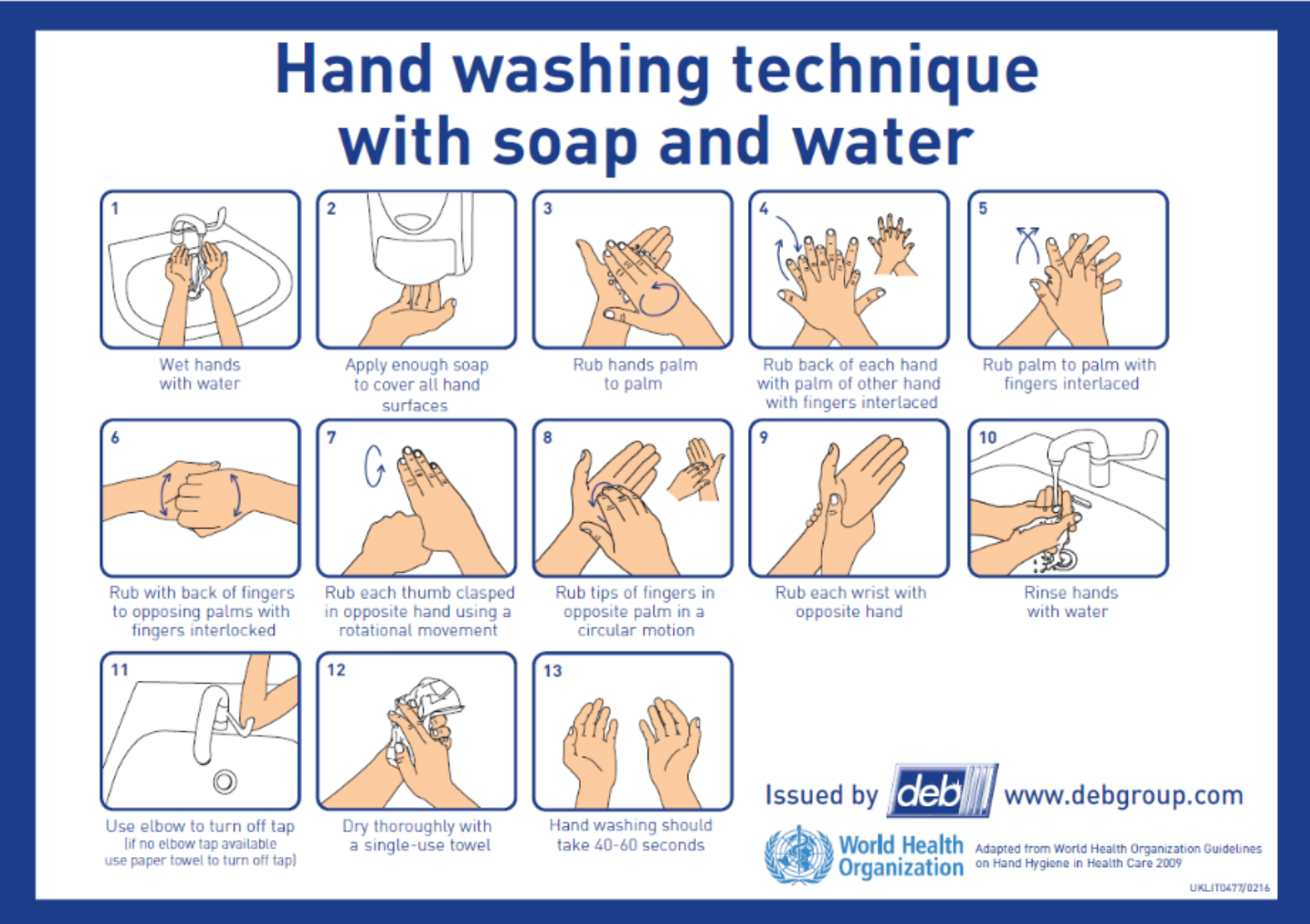}
\caption{The WHO Handwashing Steps.}
\label{fig:who_step}
\end{figure}

The quality of handwashing is strongly correlated with food safety and the general public welfare. According to the World Health Organization (WHO), food-borne illnesses sickens millions of people and is responsible for more than 400,000 deaths every year \cite{who_food}. Therefore, ensuring proper handwashing and hand hygiene of food-handlers is critical for improving food safety and reducing the spread of food-borne illnesses. The WHO has proposed a series of handwashing actions that will thoroughly clean all surfaces of the hands, as shown in Figure~\ref{fig:who_step}. In this paper, we follow the WHO guideline with an emphasis on steps 3-9 because they are the more important and challenging rubbing actions.

In our previous work\cite{ju2024mipr}, we generated a synthetic dataset to introduce controlled distribution shift for measuring the impact of hand pose and shadow on handwashing recognition. In particular, we demonstrated that both changes in hand poses and shadow can create system breakdown points and that heavy shadow can shift the breakdown points earlier. A breakdown point is where system performance starts to degrade sharply and become unreliable. Also, we explored simple mitigation strategies to the pose-induced breakdown points but did not consider mitigating the shadow-induced breakdown points. Therefore, in this paper, we explore mitigation strategies to the shadow-induced breakdown points using our synthetic dataset. Existing research has shown that learning human pose models from synthetic data can be an effective method for improving human action recognition performance on real datasets \cite{liu2019learning}. Therefore, we also exploit insights gained from synthetic data to improve system performance on real data. 

We collected two real-world outdoor handwashing datasets in \cite{ju2023robust}: the Portable51 and Farm23 dataset. Both datasets contain varied shadow conditions and particularly Farm23 contains constant and heavy outdoor shadow. Based on experimental results in \cite{ju2023robust}, we have demonstrated that system performance drops significantly across datasets due to changes in environments. Therefore, we transfer and apply the insights gained from synthetic data to real data by proposing a shadow augmentation method that will be applied while training a classifier. More specifically, we focus on evaluating model robustness on completely unseen and out-of-distribution (OoD) datasets with shadow, as model robustness is one of the most crucial factors for a real-world vision system. 

The main contributions of this paper is twofold. First, we explore and demonstrate effective strategies for mitigating the shadow-induced breakdown points using synthetic data. Then, we propose a data augmentation method called shadow augmentation to improve model robustness against outdoor shadow, and we evaluate its effectiveness using different neural network architectures and datasets. 

\section{Related Work}

\subsection{Systems for Handwashing Action Recognition}
Handwashing monitoring or action recognition has been applied for both healthcare and food-safety-related systems. In healthcare settings, a camera-based tutorial system for healthcare workers has been developed and studied by \cite{lacey_system1, lacey_system2, lacey_system3}. More recently, many deep-learning based handwashing systems have been developed. For example, a deep-learning-based handwashing quality assessment system was proposed by \cite{wang2022handwashing}. Additionally, a neural network combined with self-attention blocks was applied for classifying the rubbing actions for handwashing in healthcare \cite{chinese_journal}. Although these works have considered the WHO handwashing guidelines, they have only considered indoor medical applications, whereas we focus on outdoor and food-related handwashing settings that contain challenging shadow conditions. 

Besides systems developed for healthcare, handwashing recognition systems have also been developed for improving food-safety. A two stream network that combines spatial and temporal information has been proposed by \cite{zhong2019hand} to classify egocentric handwashing actions following the WHO steps. Furthermore, a two-stage system \cite{zhong2020multi} was developed by incorporating a coarse-to-fine recognition strategy and motion histogram images. In addition, an effective recognition system that incorporates a variety of modalities was proposed for cross-domain handwashing action recognition \cite{zhong2021designing}. These systems have shown promising results for recognizing handwashing actions under different environments. However, they did not consider the more challenging shadow-heavy environments or consider how shadow can impact system performance.  

\subsection{Shadow in Computer Vision and Video Systems}
Since shadow is inevitable in the outdoors, many existing works have studied the problem of shadow detection and shadow removal. To build a robust video surveillance system, a shadow elimination method using illumination images was proposed by \cite{matsushita2002shadow}. Furthermore, a large-scale synthetic dataset was created and used to improve the performance of methods for shadow removal \cite{inoue2020learning}. Also, shadow has been shown to mislead vehicle detection and tracking systems \cite{song2014vehicle}. A shadow detection method for analyzing traffic surveillance videos was proposed by \cite{shi2020new}. These works have all discussed the negative impact of shadow on video systems. In addition, the solutions to mitigate the impact of shadow are associated with detecting and removing shadow from the scene. However, in this paper, we explore whether we can improve system robustness by adding shadow to the model training process. 

\section{Datasets}

\subsection{Handwashing Actions}
As discussed earlier in Section~\ref{sec:intro}, we only focus on the 7 rubbing actions provided by the WHO. For simplicity, we shorten the names of these actions to be: rub back, rub palm, rub fingers interlaced, rub thumb, rub tips, rub back fingers, and rub wrist. 

\subsection{Synthetic Dataset}
\begin{figure}[htb!]
\centering
\begin{subfigure}[t]{0.31\linewidth}
    \includegraphics[width=\textwidth]{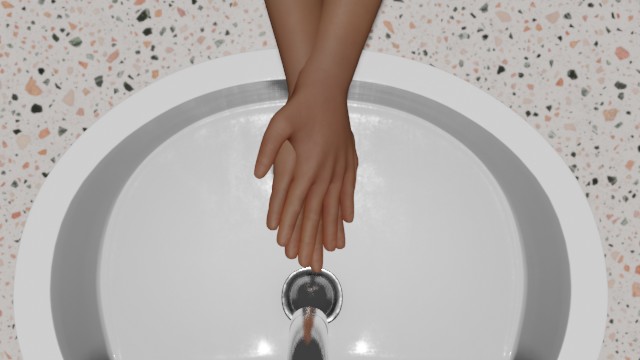}
    \caption{Rub back $-45^{\circ}$}
\end{subfigure}%
\hspace{.1cm}
\begin{subfigure}[t]{0.31\linewidth}
    \includegraphics[width=\textwidth]{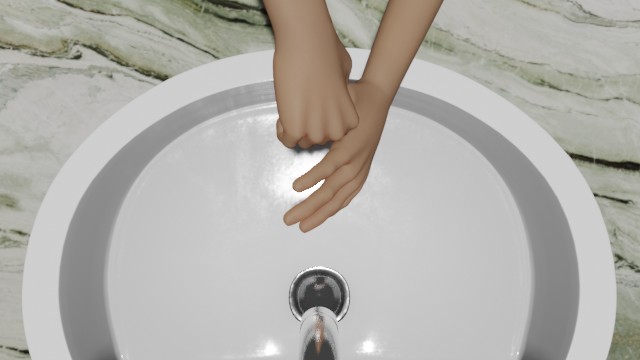}
    \caption{Rub thumb $-45^{\circ}$}
\end{subfigure}%
\hspace{.1cm}
\begin{subfigure}[t]{0.31\linewidth}
    \includegraphics[width=\textwidth]{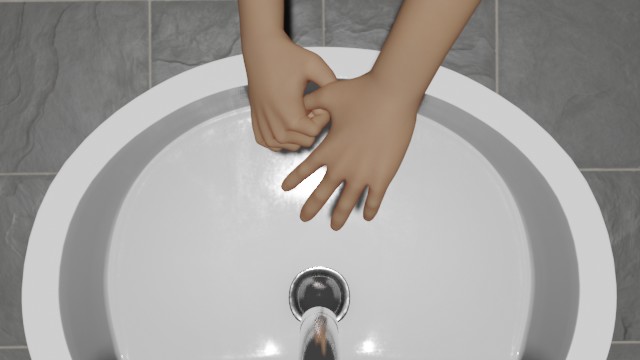}
    \caption{Rub thumb $45^{\circ}$}
\end{subfigure}%
\caption{Examples of different hand poses.}
\label{fig:action_ex}
\end{figure}

The synthetic dataset is the same as introduced in \cite{ju2024mipr}. Synthetic images of handwashing are generated with different skin tones and background textures using Blender. In addition, hand poses from -90 to 90 degrees of rotation are generated for each action. For example, Figure~\ref{fig:action_ex} shows sample images of different actions and hand poses. In addition, images with different shadow properties are generated including different shadow sizes, shadow intensities, and shadow placements. Shadow is generated by placing a cylindrical pole object between the light source and the hands. There are 4 different shadow intensity levels which are characterized by the alpha value of the pole object: alpha02, alpha04, alpha06, and alpha08. Higher alpha values correspond to heavier shadow. In addition, there are 3 different shadow sizes which are controlled by the width of the pole: pole05 (smallest), pole10, and pole15 (largest). 

\begin{figure}[htb!]
\centering

\begin{subfigure}[t]{0.31\linewidth}
    \includegraphics[width=\textwidth]{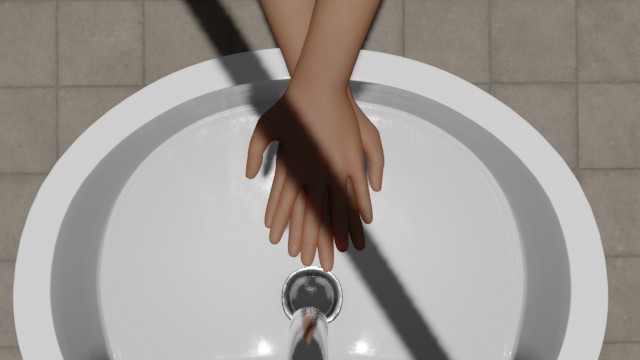}
    \caption{Shadow size 1}
\end{subfigure}%
\hspace{.1cm}
\begin{subfigure}[t]{0.31\linewidth}
    \includegraphics[width=\textwidth]{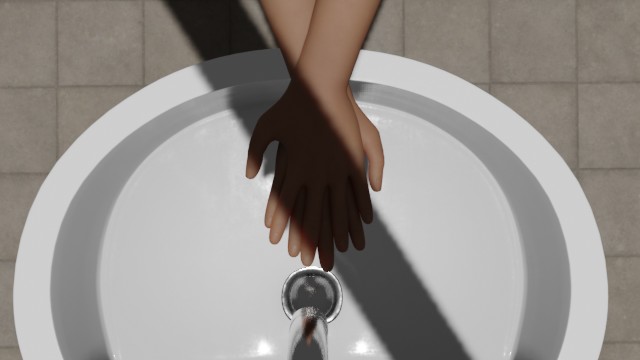}
    \caption{Shadow size 2}
\end{subfigure}%
\hspace{.1cm}
\begin{subfigure}[t]{0.31\linewidth}
    \includegraphics[width=\textwidth]{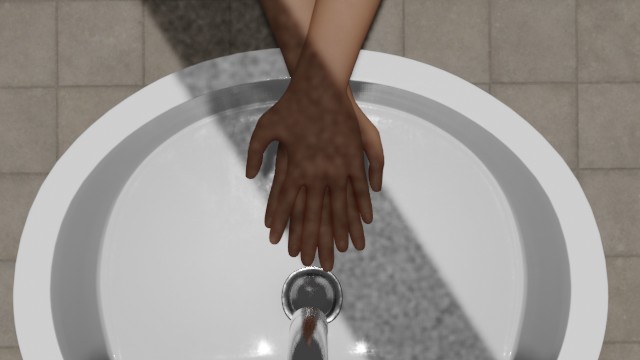}
    \caption{Shadow intensity}
\end{subfigure}%

\caption{Examples of images with different shadow attributes.}
\label{fig:shadow_ex}
\end{figure}

Figure~\ref{fig:shadow_ex} shows sample images of different shadow sizes and intensities. This dataset contains varied hand poses for 5 of the 7 rubbing actions, excluding rub wrist and rub back fingers. However, shadow images were only generated for the actions of rub back and rub thumb due to computational limits.  

\subsection{Real-world Datasets}
\begin{figure}[htb!]
\centering

\begin{subfigure}[t]{0.31\linewidth}
    \includegraphics[width=\textwidth]{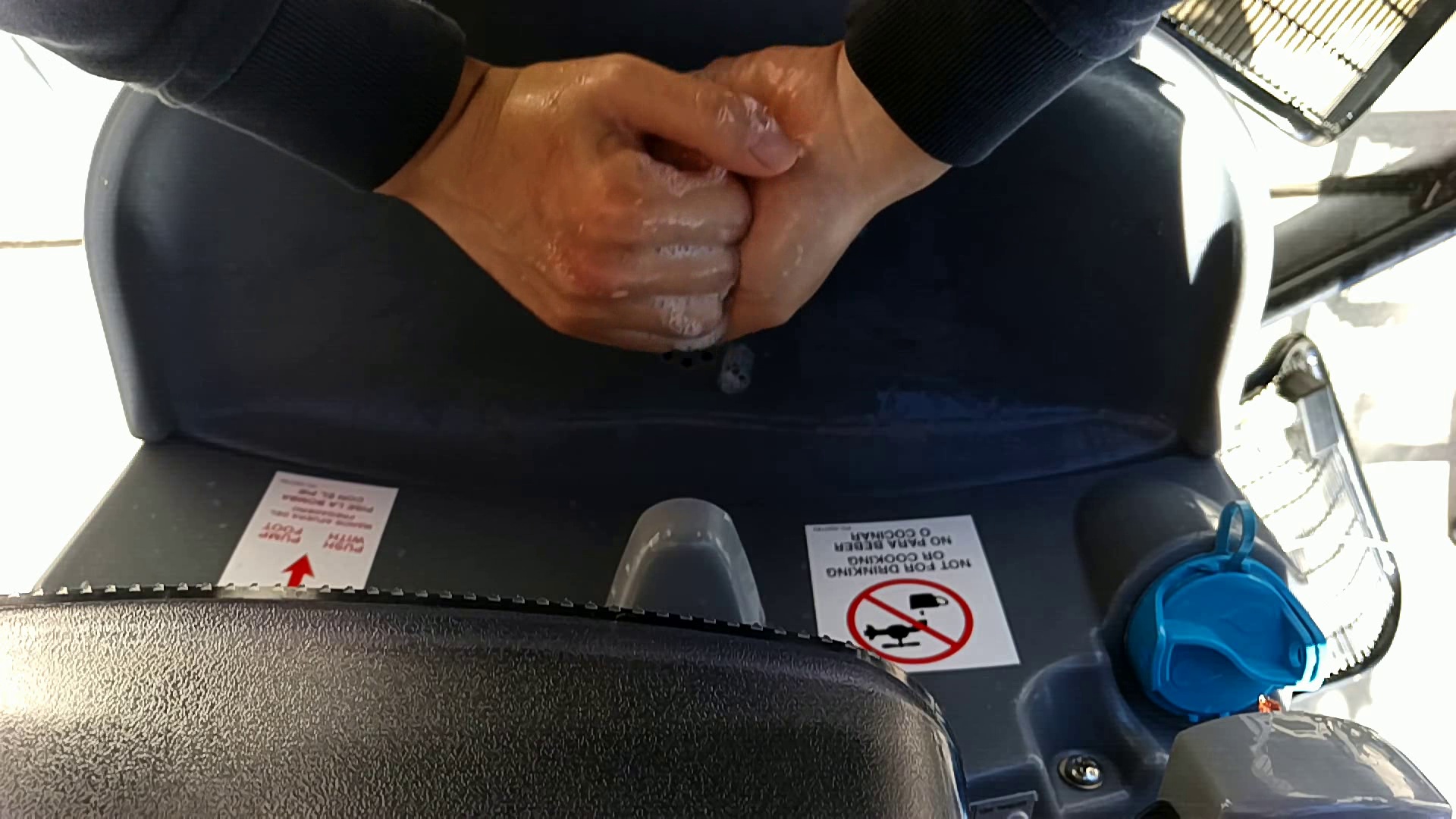}
    \caption{Portable51}
\end{subfigure}%
\hspace{.1cm}
\begin{subfigure}[t]{0.31\linewidth}
    \includegraphics[width=\textwidth]{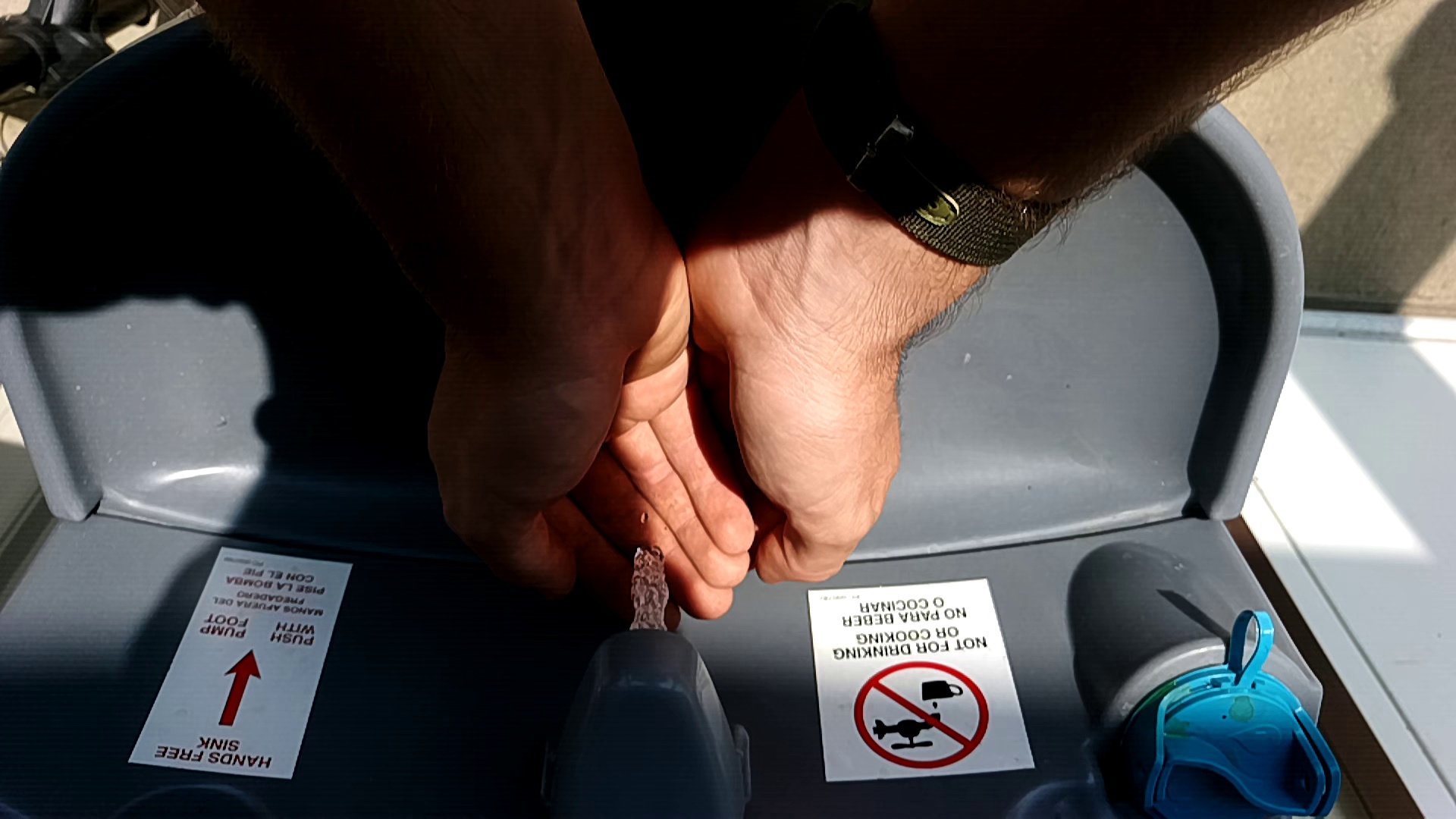}
    \caption{Farm23}
\end{subfigure}%
\hspace{.1cm}
\begin{subfigure}[t]{0.31\linewidth}
    \includegraphics[width=\textwidth]{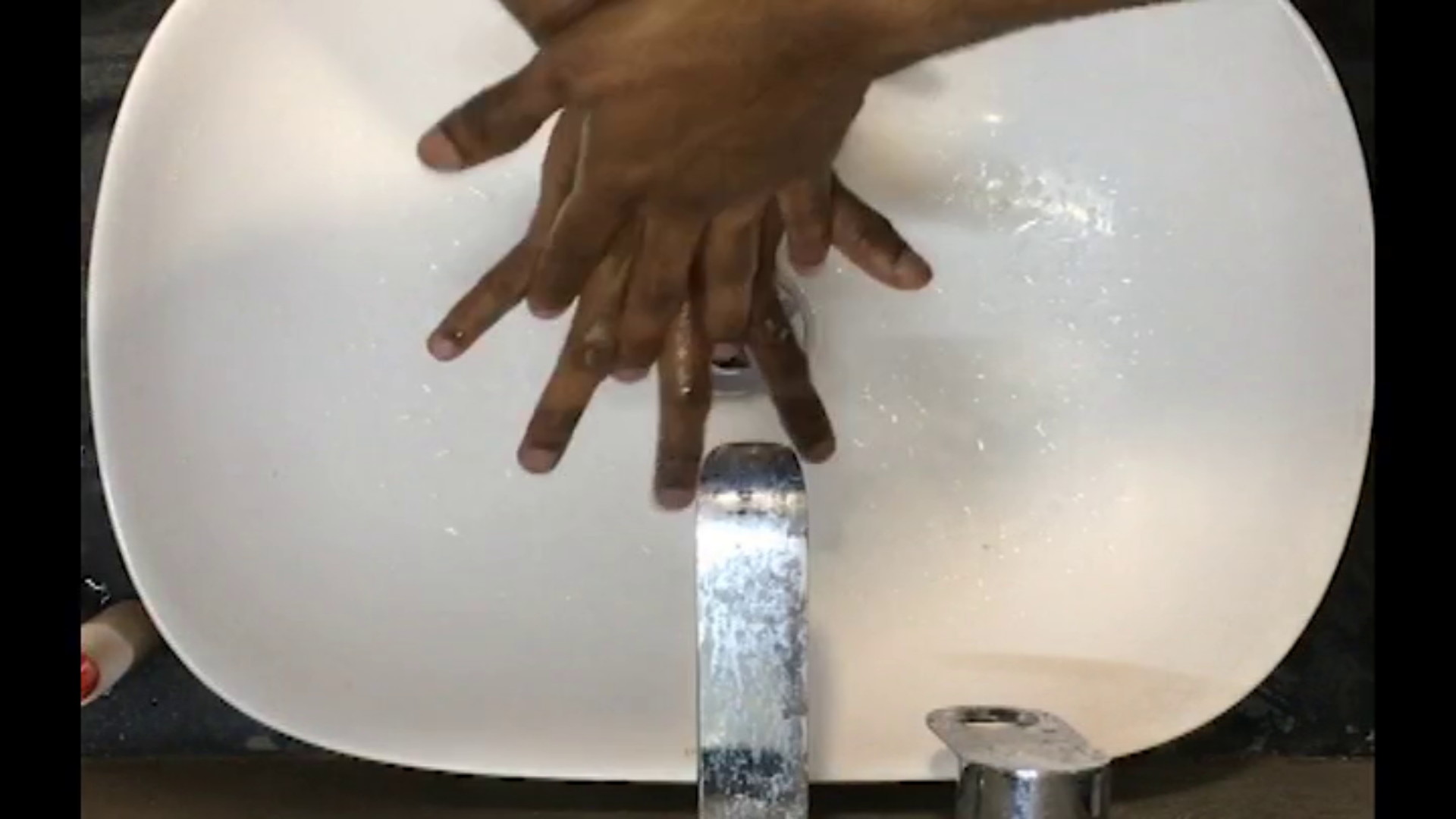}
    \caption{KHWD}
\end{subfigure}%

\caption{Examples of each dataset.}
\label{fig:dataset_ex}
\end{figure}

In this paper, we also use 3 real-world datasets with unique attributes. More specifically, we use the Portable51 and Farm23 datasets collected in \cite{ju2023robust} and the public Kaggle Hand Wash Dataset (KHWD)\cite{kaggle}. All 3 datasets contain videos of the 7 rubbing actions following the WHO guideline. Portable51 contains videos from 51 unique participants and was captured both indoors and outdoors with a moderate amount of outdoor shadow. Farm23 contains videos from 12 new participants with no overlap between the ones in Portable51. These two datasets are captured with a portable sink and diverse participants with different ages and skin tones to better represent real-world deployment scenarios. 
In addition, Farm23 is captured entirely outdoors and contain challenging scenarios with constant and heavy shadow. Lastly, the KHWD contains videos of indoor handwashing with limited presence of shadow and skin-tone variation. Figure~\ref{fig:dataset_ex} shows one sample image for each dataset. As seen from the figures, Farm23 contains much heavier shadow compared to the other datasets. Portable51 contains moderate shadow while KHWD contains little to no shadow.

\vspace{-.5cm}
\subsection{Shadow Augmentation}
\label{data:shadow_aug}
\begin{figure}[htb!]
\centering

\begin{subfigure}[t]{0.45\linewidth}
    \includegraphics[width=\textwidth]{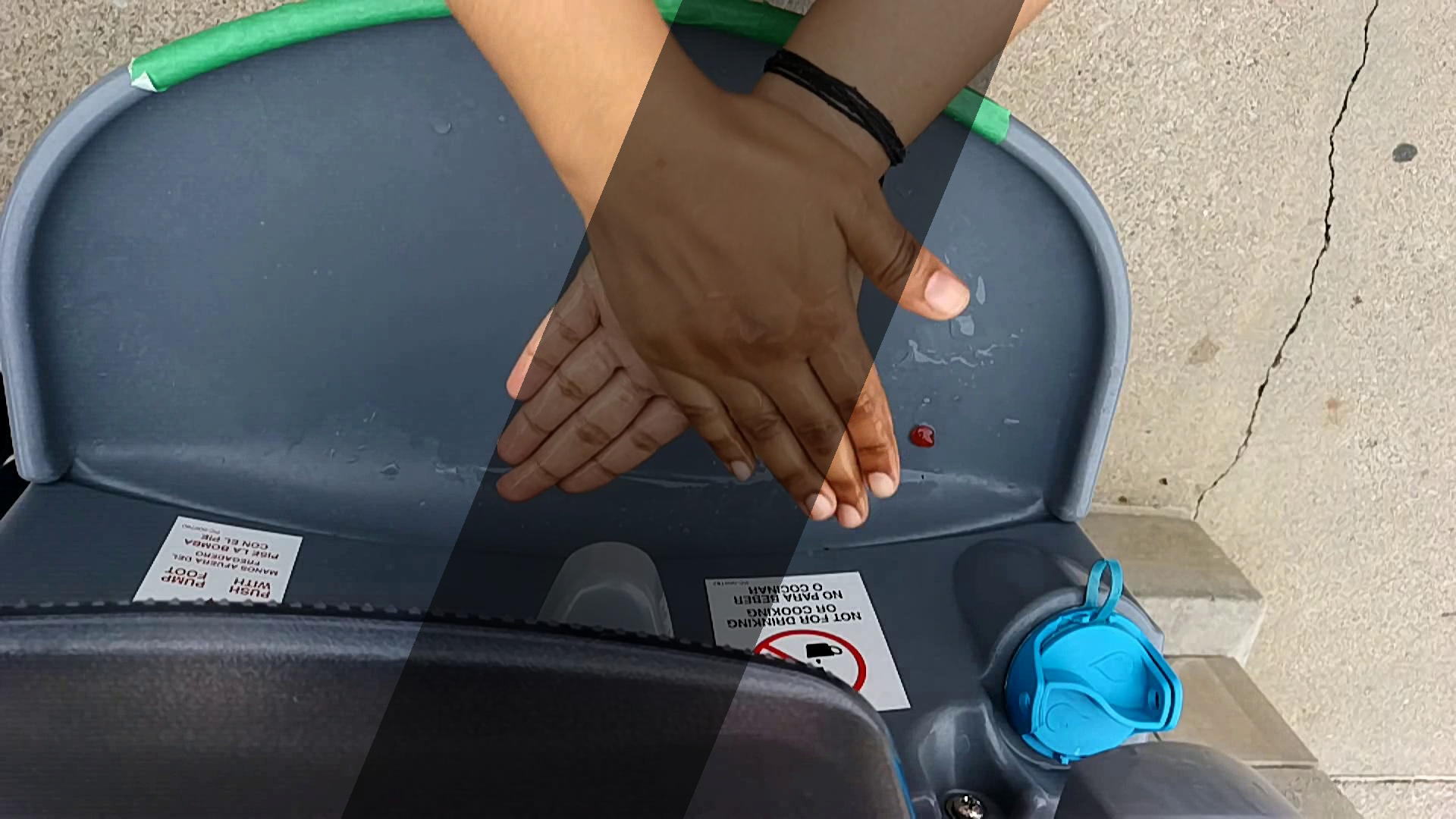}
    \caption{Portable51 + shadow ex.1}
    \label{subfig: p51_shadow1}
\end{subfigure}%
\hspace{.1cm}
\begin{subfigure}[t]{0.45\linewidth}
    \includegraphics[width=\textwidth]{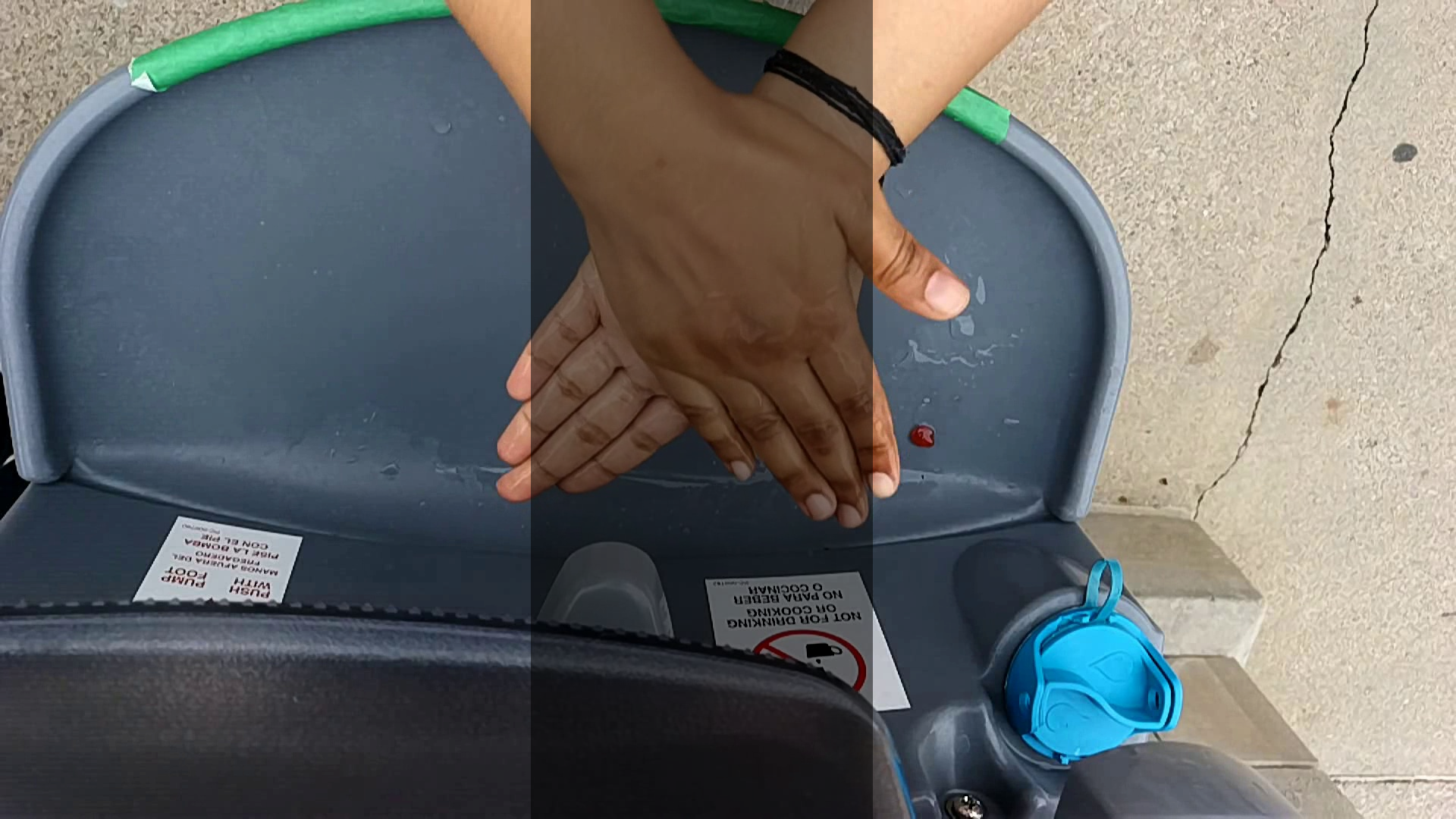}
    \caption{Portable51 + shadow ex.2}
    \label{subfig: p51_shadow2}
\end{subfigure}
\vspace{.1cm}
\begin{subfigure}[t]{0.45\linewidth}
    \includegraphics[width=\textwidth]{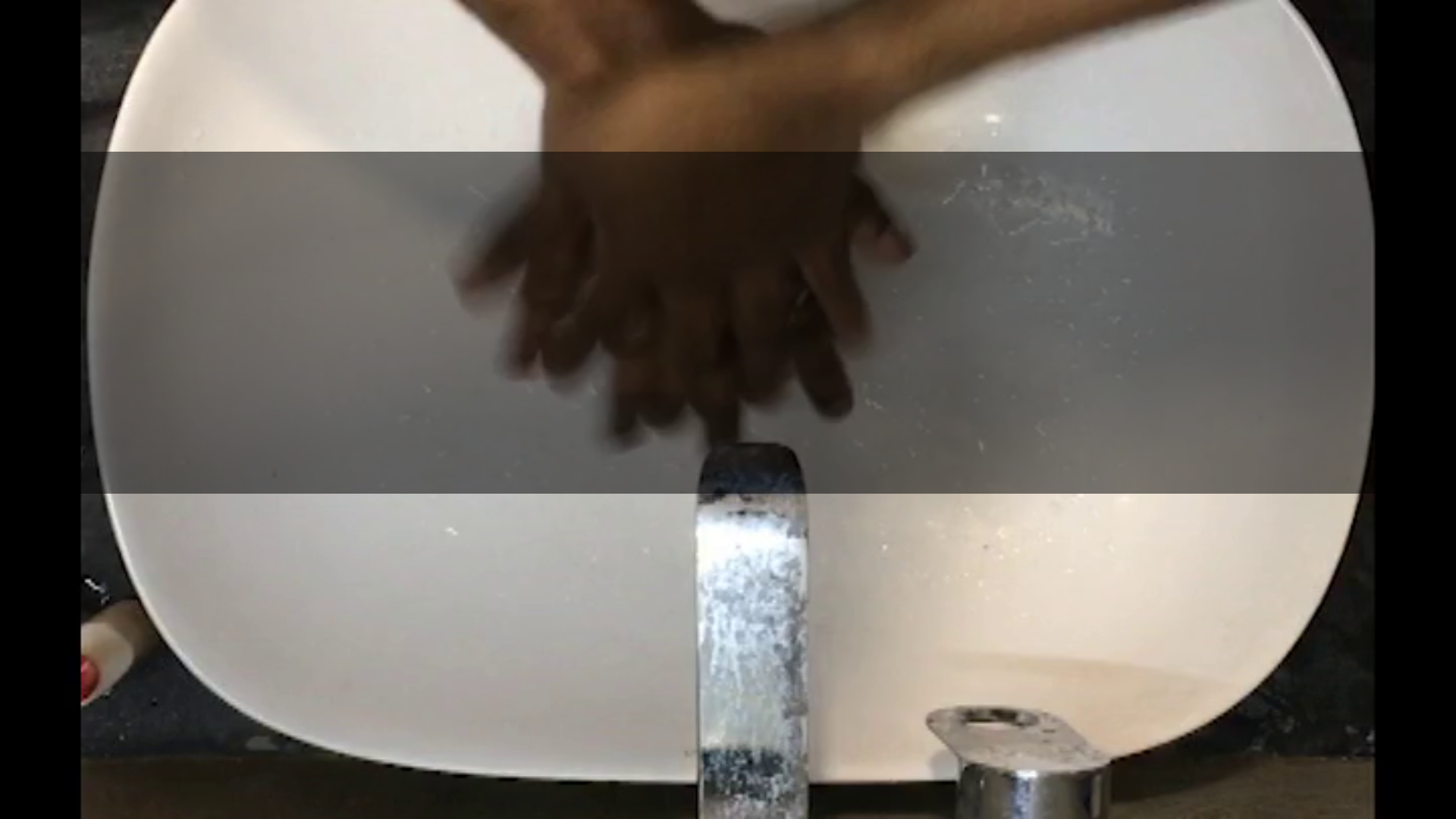}
    \caption{KHWD + shadow ex.1}
\end{subfigure}%
\hspace{.1cm}
\begin{subfigure}[t]{0.45\linewidth}
    \includegraphics[width=\textwidth]{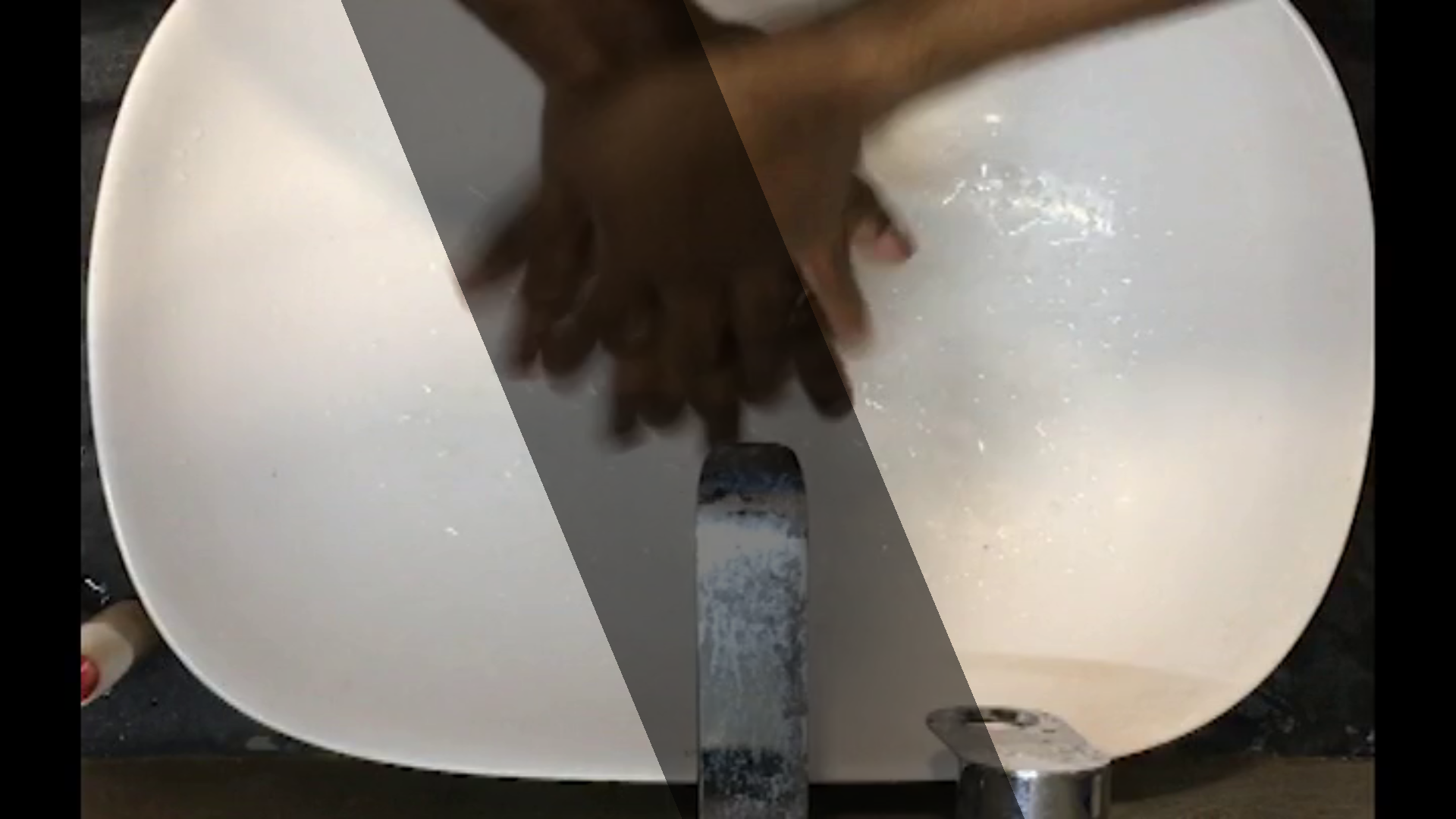}
    \caption{KHWD + shadow ex.2}
\end{subfigure}%

\caption{Examples of shadow augmentation on the Portable51 and KHWD dataset.}
\label{fig:p51_shadow}
\end{figure}

In this section, we discuss the process of shadow augmentation, which adds synthetic shadow to real-world images. The motivation for adding shadow to real images is to explore whether the added shadow can help classifiers become more robust to shadowy test conditions if used as training data. We follow a similar but simplified strategy for adding shadow compared to the synthetic dataset. First, we define a set of 4 vertices to draw the polygon that will contain the shadow. The vertices are fully adjustable so that the shape of shadow can also be adjusted. 
Next, we multiply the alpha channel of the pixels within the polygon by a controllable shadow factor (s.f.), which is a constant less than 1, to adjust shadow intensity. For our experiments, we use 4 different polygons for shadow which represent different orientations and placements of natural shadow, and we fix shadow intensity for better reproducibility. This process of shadow augmentation can be applied to any dataset. 
Figure~\ref{fig:p51_shadow} shows sample images of shadow augmentation applied to the Portable51 and KHWD dataset, using a s.f. of 0.5. Also, the different orientations and placements can be seen from the figure. 

\vspace{-.3cm}
\section{Experiments and Results}

\subsection{Experimental Setup for Synthetic Data}
As seen from \cite{ju2024mipr} using our synthetic dataset, shadow causes a handwashing recognition system's performance to drop significantly. In particular, as shadow intensity increases from 2 to 8, the average top1 accuracy decreases by 57\%. Also, as shadow size increases from 1 to 3, the average accuracy decreases by more than 12\%.
Here, we explore strategies to mitigate the breakdown points and the degradation in performance caused by shadow-induced distribution shift. In particular, we explore which shadow size and intensity, characterized by pole width and pole transparency values respectively, is the best for mitigating the breakdown points. To accomplish this, we keep the 0-degree and shadow-free data as fixed training data. Then, we train a separate model with each of the following sets of additional data with a specific shadow attribute: alpha02, alpha04, alpha06, alpha08, pole05, pole10, and pole15. Images for each shadow attribute are randomly and uniformly sampled across the other shadow attributes. For example, images in the additional training set for alpha02 are sampled across images that contain pole width values from 5 to 15, 2 different translations, and 2 different rotations. Images are also sampled from all background and skin tones. In addition, to control the type of distribution shift added to the training set, we only sample from the 0-degree hand pose for all actions. 
Also, since we have only generated images with shadow for the actions of rub thumb and rub back, additional shadowy training images are only sampled from these two actions. Furthermore, the trained models are evaluated on these two actions and 3 shadow attributes: shadow size (simulated by pole width), shadow intensity (simulated by pole transparency/alpha), and shadow placements (simulated by pole placements).

In total, 3,200 additional images are added for exploring the optimal shadow size and 2,400 additional images are added for exploring the optimal shadow intensity. These images are added to the original 5,000 shadow-free images at the 0-degree hand pose. Due to the constraints on datasets and computational limits discussed in \cite{ju2024mipr}, we finetune a lightweight MobileNetV3\cite{mbnetv3} that is pretrained on ImageNet \cite{imagenet}. 

\subsection{Which Shadow Properties are Better for Mitigating the Breakdown Points?}
\begin{figure}[htb!]
\centering
\begin{subfigure}[t]{0.47\linewidth}
    \includegraphics[width=\textwidth]{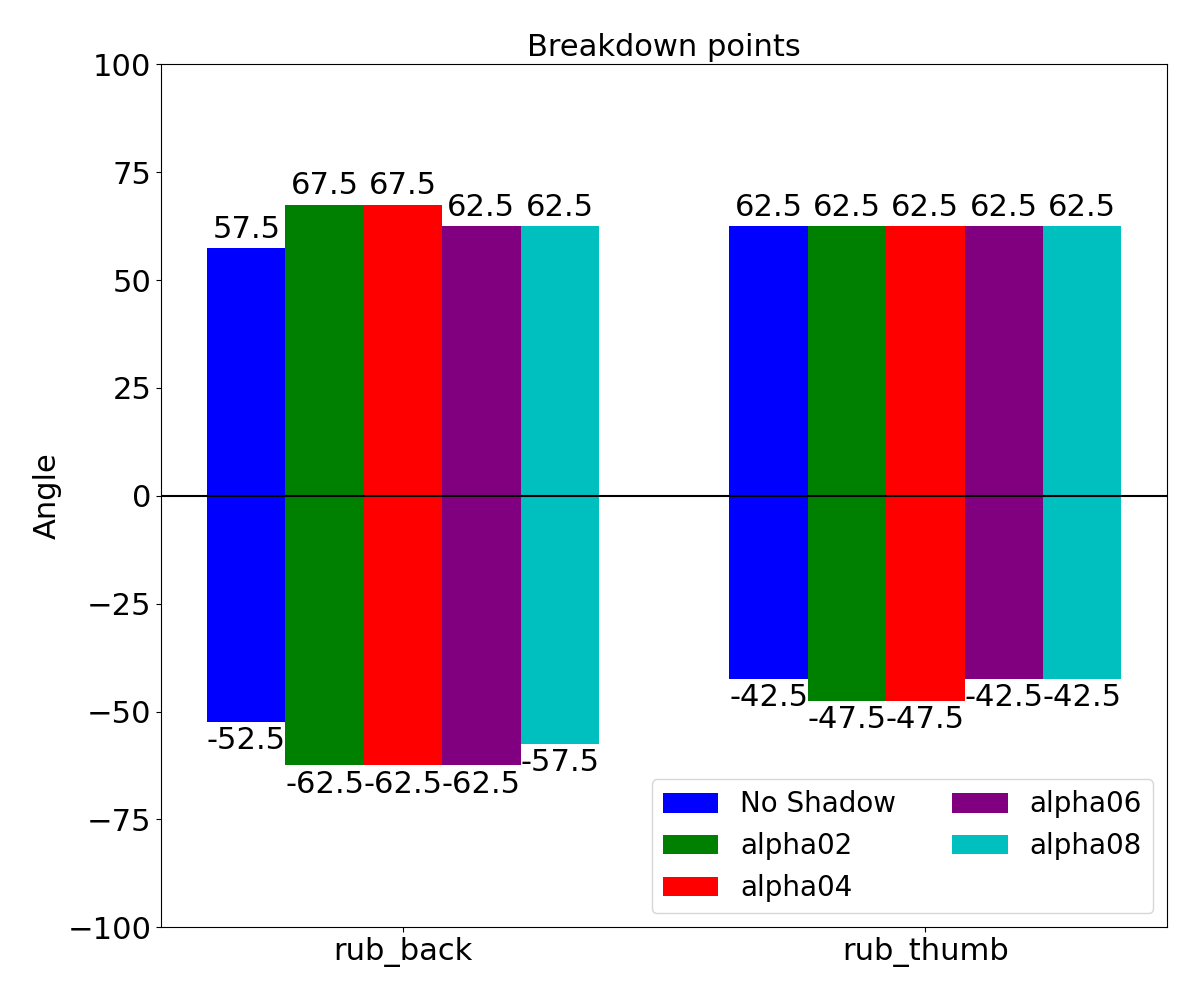}
    \caption{Breakdown points of adding lighter shadow (alpha02).}
\end{subfigure}%
\hspace{.1cm}
\begin{subfigure}[t]{0.47\linewidth}
    \includegraphics[width=\textwidth]{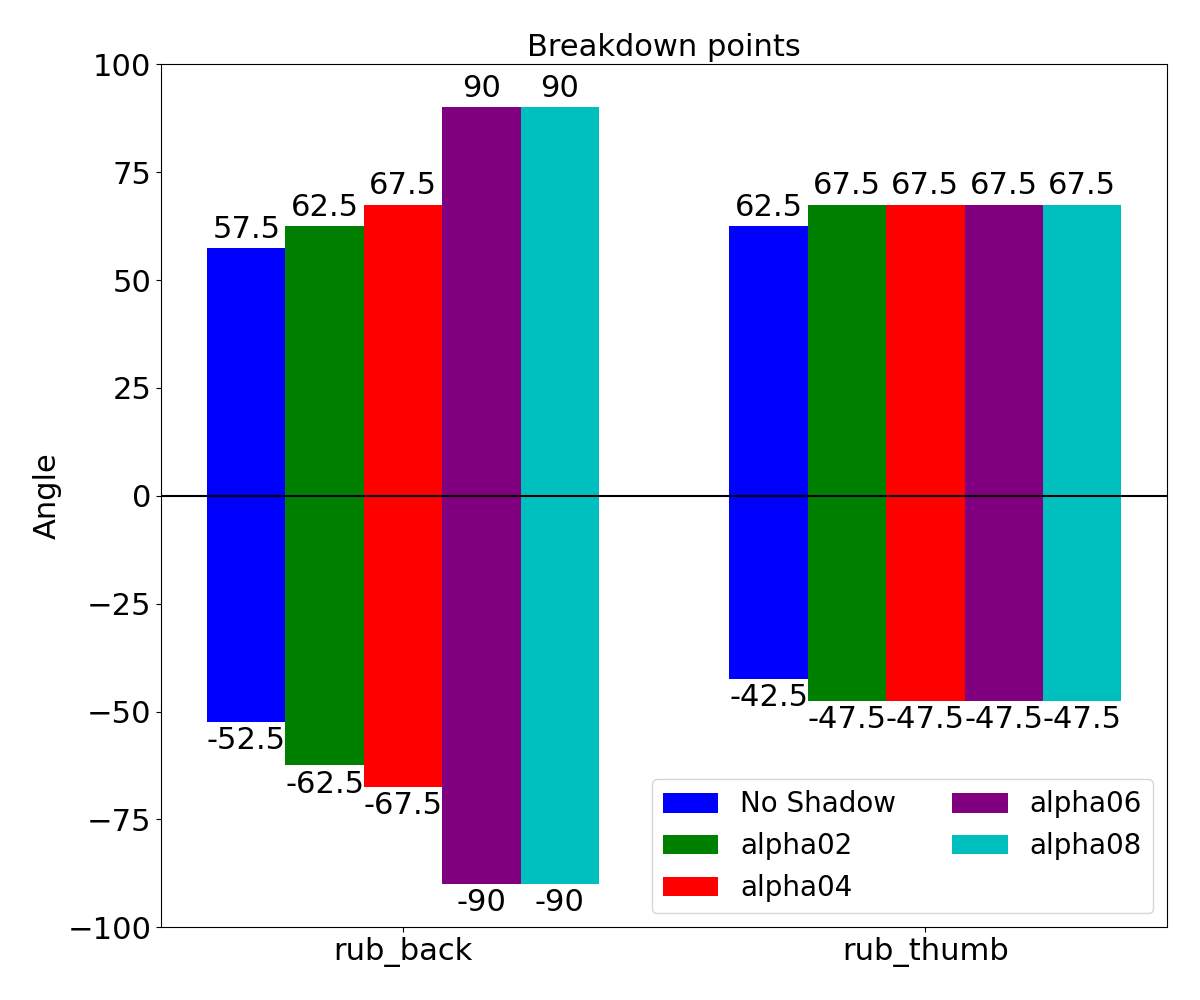}
    \caption{Breakdown points of adding heavier shadow (alpha08).}
\end{subfigure}%
\caption{Breakdown points using two different sets of added training data with different shadow intensities.}
\label{fig:shadow_alpha_add_bdpts}
\end{figure}

We use top1 accuracy and the breakdown points\cite{ju2024mipr} for evaluation. A breakdown point occurs when top1 accuracy drops below 60\% or when top1 accuracy drops by more than 15\% for a consecutive 5-degree change in hand pose. First, we demonstrate results for exploring which shadow intensity is better. We omit the top1 accuracy plots and directly compare the breakdown points. Also, we only show two sample results in this section. 
Figure~\ref{fig:shadow_alpha_add_bdpts} shows the breakdown points of two models trained with different shadow intensity and tested across varying shadow intensity. It can be observed that the breakdown points of the model trained with alpha08 have significantly higher absolute values, which indicates that the breakdown points are shifted to much later. It is worth mentioning that a breakdown point of $\pm90$ degree indicates there is no breakdown point. Another important finding is that the breakdown points of rub thumb shifted later but only slightly. This could be due to the unique pose of rub thumb. 



\begin{figure}[htb!]
\centering
\begin{subfigure}[t]{0.47\linewidth}
    \includegraphics[width=\textwidth]{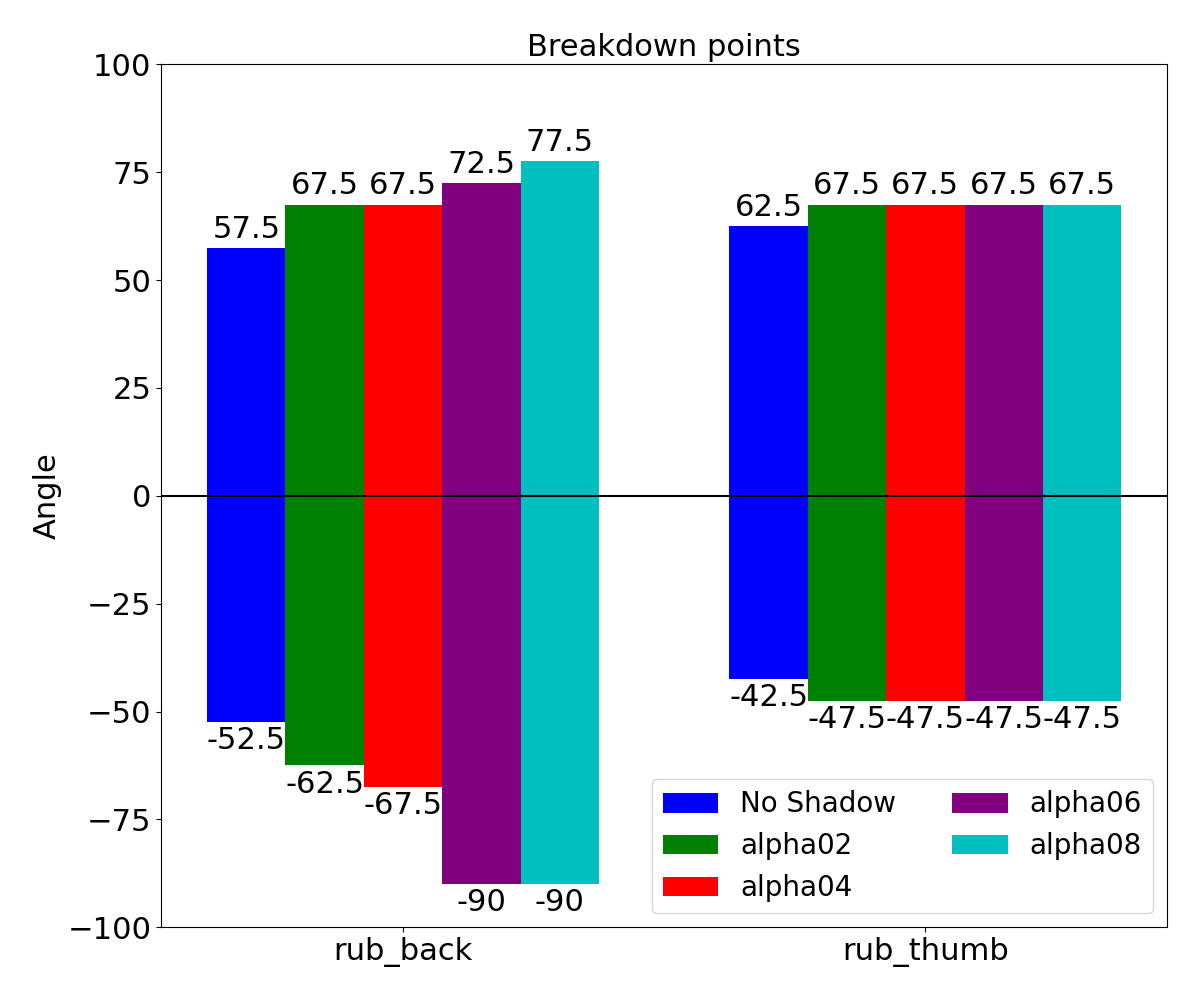}
    \caption{Breakdown points of adding smaller shadow (pole05).}
\end{subfigure}%
\hspace{.1cm}
\begin{subfigure}[t]{0.47\linewidth}
    \includegraphics[width=\textwidth]{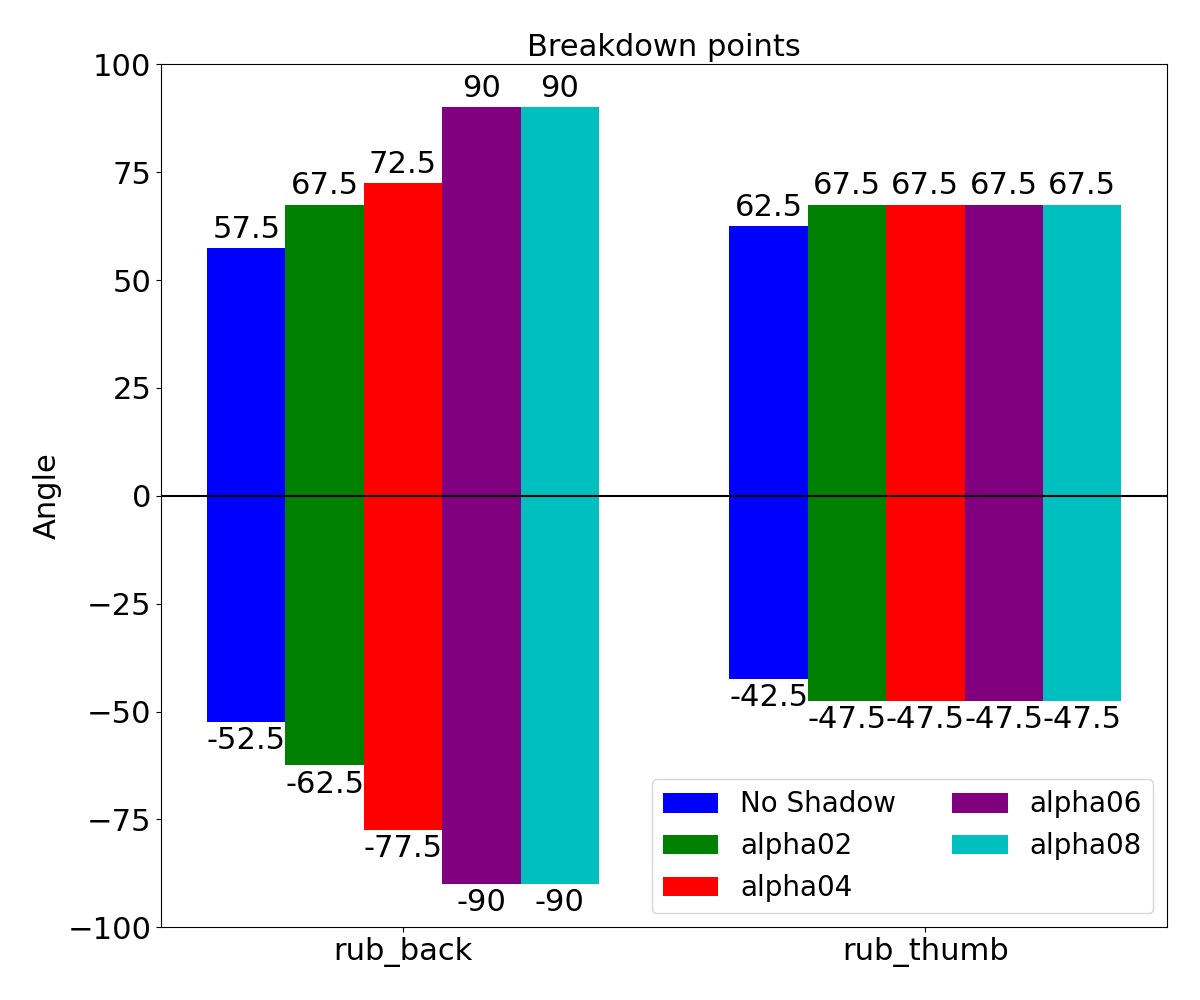}
    \caption{Breakdown points of adding larger shadow (pole15).}
\end{subfigure}%
\caption{Breakdown points using two different sets of added training data with different shadow sizes.}
\label{fig:shadow_pole_add_bdpts}
\end{figure}

Next, we demonstrate the results of training with additional data containing different shadow sizes. Again, we omit the top1 accuracy plots. 
Figure~\ref{fig:shadow_pole_add_bdpts} shows the breakdown points of adding shadow data with different pole widths. As can be seen from the plots, training with pole15 shadow data results in later breakdown points. Also, it can be observed that the amount of shift is not as significant when compared to training with different pole transparency. Furthermore, the breakdown points for rub thumb did not change across the two models and test shadow conditions. This shows that rub thumb benefits less from additional training data.



Table~\ref{tab:add_shadow} shows the complete results of both the positive and negative breakdown points after training with each of the different shadow properties. For each type of shadow, results are averaged across all testing shadow conditions including shadow size, shadow intensity, and shadow placement. From the table, it can be observed that the alpha06 and alpha08 resulted in better breakdown points for negative and positive hand rotational angles, respectively. This indicates that heavier shadow is more beneficial for mitigating the shadow-induced breakdown points. From the last 3 rows of the table, it can be concluded that pole15 shadow is the best additional training data compared to other pole widths. This indicates that larger/wider shadow is more optimal for mitigating the shadow-induced breakdown points. Overall, to improve the breakdown points and the significant performance degradation caused by shadow, it is better to add training data with shadow that is more intense and larger. 

\begin{table}[htb!]
\centering
\begin{tabular}{|c|c|c|}
\hline
\begin{tabular}[c]{@{}c@{}}Added Training \\ Shadow Type\end{tabular} & \begin{tabular}[c]{@{}c@{}}Avg. Breakdown \\ Points (+)\end{tabular} & \begin{tabular}[c]{@{}c@{}}Avg. Breakdown \\ Points (-)\end{tabular} \\ \hline
alpha02                    & $63.41^{\circ}$                        & $53.18^{\circ}$   \\ \hline
alpha04                    & $66.82^{\circ}$                        & $57.05^{\circ}$   \\ \hline
alpha06                    & $68.75^{\circ}$                        & $\mathbf{62.16^{\circ}}$
\\ \hline
alpha08                    & $\mathbf{69.32^{\circ}}$               & $61.82^{\circ}$   \\ \hline
pole05                     & $68.18^{\circ}$                        & $60.23^{\circ}$    \\ \hline
pole10                     & $69.89^{\circ}$                        & $62.84^{\circ}$    \\ \hline
pole15                     & $\mathbf{70.68^{\circ}}$             & $\mathbf{63.98^{\circ}}$              \\ \hline
\end{tabular}
\caption{Average breakdown points for different types of added training shadow.}
\label{tab:add_shadow}
\end{table}

\subsection{Experimental Methodology for Real Datasets and Shadow Augmentation}
We design 3 phases of experiments to demonstrate the effectiveness of shadow augmentation for real data. To begin with, we design the first phase of experiments to compare the performance of shadow augmentation with other data augmentation methods that adjust image brightness. For all models, training data is fixed to be images from the Portable51 train data. We train individual models that contain the following data augmentation steps: shadow augmentation, reducing the overall brightness by 50\%, reducing the overall brightness by 50\% with a probability of 0.5, and random color jittering that adjusts image brightness between 0.25 and 1. A common data augmentation step of random horizontal flip is applied for all experiments. Also, shadow augmentation is only applied to 50\% of all training images. All data augmentation is only applied to the training data. 
For evaluation, we use the Portable51 test set as the in-distribution set and the entire Farm23 and KHWD datasets as the out-of-distribution (OoD) evaluation data for model robustness. Classification performance on the Farm23 dataset is of particular interest and significance because it contains constant and heavy shadow. Therefore, evaluation results on Farm23 will be indicative of a system's performance against heavy shadow conditions. We finetune ResNet50\cite{resnet} models pretrained on ImageNet\cite{imagenet}. 

The second phase of experiments is to verify the effectiveness of shadow augmentation using different neural networks. In this phase, we experiment with a variety of neural networks including ResNet50, ResNet152, and ViT\cite{vit}. In addition, we keep the training set fixed to be the original Portable51 train set. Then we train separate models for each neural network architecture by applying shadow augmentation. The evaluation datasets remain the same as in the first phase. We observe whether there exists a consistent trend for performance across different architectures.

In addition to experimenting with different NN architectures, we experiment using different training datasets as the third phase of experiments. We apply shadow augmentation to the KHWD train set and compare its performance with its baseline model. The KHWD dataset was captured entirely indoors without outdoor shadow instances. Therefore, testing data with outdoor shadow can be challenging for a model that has only been trained with KHWD data. The goal is to explore whether applying shadow augmentation to a dataset without shadow images improves the trained model's robustness towards OoD datasets with shadow. 
The evaluation datasets are the KHWD test set, the entire Portable51 dataset and Farm23 dataset, with the latter two being completely unseen and OoD to the KHWD train set. Here, we also finetune pretrained ResNet50 models.  

All hyperparameters such as the optimizer, learning rate, and number of epochs are kept the same for all experiments. Since the datasets used in this series of experiments include 7 actions, we experiment with all 7 rubbing actions to better evaluate the effectiveness of our shadow augmentation method. Furthermore, we run all experiments using 3 different random seeds and average results. 

\subsection{Results and Comparison with Other Data Augmentation Methods}
Because we want to explore the general effectiveness of shadow augmentation, we report only the overall top1 accuracy across all actions instead of listing results for individual actions. Table~\ref{tab:data_aug_compare} shows the overall top1 accuracy for the baseline ResNet50 model and models trained with different data augmentation methods. As can be seen from the results, applying shadow augmentation to 50\% of the training images improves performance on both the in-distribution test set and the completely OoD test sets compared to other tested methods. The improvement from shadow augmentation when compared to other methods is relatively marginal on the Portable51 test set. However, model robustness against distribution shifts is noticeably improved, as demonstrated by results on the Farm23 and KHWD datasets. Particularly, performance improved by more than 8\% on the challenging Farm23 dataset with constant and heavy shadow. Among the other data augmentation methods, simply reducing the brightness of all training images by a constant factor has resulted in overfitting and much worse performance. Random color jittering proves to be effective and demonstrates similar performance compared to shadow augmentation. However, shadow augmentation still outperforms by a small margin. These results demonstrate that shadow augmentation is more effective at improving model robustness, especially against shadow conditions, than simply adjusting the brightness of the images. Next, we verify the effectiveness of shadow augmentation using different architectures and training dataset.

\begin{table}[htb!]
\centering
\begin{tabular}{|c|c|c|c|}
\hline
Method                                                          & Portable51-test & Farm23 & KHWD         \\ \hline
Baseline                                                        & 0.665          & 0.440   & 0.304       \\ \hline
Reduce brightness by 50\%                                       & 0.181          & 0.159 & 0.174          \\ \hline
\begin{tabular}[c]{@{}c@{}}Reduce brightness by 50\% \\ with 0.5 prob.\end{tabular}      & 0.669  & 0.444 & 0.298         \\ \hline
\begin{tabular}[c]{@{}c@{}}Random color jitter \\ brightness {[}0.25, 1{]}\end{tabular} & 0.671 & 0.457 & 0.326          \\ \hline
Shadow augmentation (s.f.=0.5)                             & \textbf{0.672} & \textbf{0.476} & \textbf{0.332} \\ \hline
\end{tabular}
\caption{Overall accuracy for different data augmentation methods on different evaluation datasets.}
\label{tab:data_aug_compare}
\end{table}

\subsection{Results using Different Network Architectures and Datasets}
Table~\ref{tab:nn_shadow_aug} shows the comparison of the baseline model and the model with shadow augmentation across 3 different test datasets and using 3 different neural network architectures. The baseline models in all cases have only been trained on the Portable51 train data.
As can be seen from the table, models trained with shadow augmentation demonstrate superior performance on all test datasets including the completely unseen and OoD test sets. The table also shows the percentage of increase over the baseline model. We observe that the improvement on the Portable51 test set is less significant. However, the improvements to system robustness on unseen datasets are substantial. In particular, ViT with shadow augmentation demonstrates over 10\% increase on both OoD datasets. Furthermore, ResNet152 with shadow augmentation results in more than 26\% increase when testing on the KHWD dataset. By demonstrating system performance across different evaluation datasets and NN architectures, we show effectiveness of the shadow augmentation method on improving system robustness towards distribution shift across datasets.

\begin{table*}[t]
\centering
\begin{tabular}{|c|cc|cc|cc|}
\hline
\multirow{2}{*}{\begin{tabular}[c]{@{}c@{}}NN \\ Architecture \end{tabular}} & \multicolumn{2}{c|}{Portable51-test}                           & \multicolumn{2}{c|}{Farm23}                                 & \multicolumn{2}{c|}{KHWD}                              \\ \cline{2-7} 
                                 & \multicolumn{1}{c|}{Baseline} & Shadow Aug.             & \multicolumn{1}{c|}{Baseline} & Shadow Aug.              & \multicolumn{1}{c|}{Baseline} & Shadow Aug.              \\ \hline
ResNet50                         & \multicolumn{1}{c|}{0.665}    & \textbf{0.672 (+1.1\%)} & \multicolumn{1}{c|}{0.440}     & \textbf{0.476 (+8.2\%)}  & \multicolumn{1}{c|}{0.304}    & \textbf{0.332 (+9.2\%)}  \\ \hline
ResNet152                        & \multicolumn{1}{c|}{0.696}    & \textbf{0.701 (+0.7\%)} & \multicolumn{1}{c|}{0.436}    & \textbf{0.477 (+9.4\%)}  & \multicolumn{1}{c|}{0.377}    & \textbf{0.478 (+26.8\%)} \\ \hline
ViT                              & \multicolumn{1}{c|}{0.695}    & \textbf{0.704 (+1.3\%)} & \multicolumn{1}{c|}{0.393}    & \textbf{0.463 (+17.8\%)} & \multicolumn{1}{c|}{0.532}    & \textbf{0.590 (+10.9\%)}  \\ \hline
\end{tabular}
\caption{Overall accuracy for applying shadow augmentation using different evaluation datasets and different neural network architectures.}
\label{tab:nn_shadow_aug}
\end{table*}

In addition, we investigate whether shadow augmentation can generalize to other training datasets besides the Portable51 data. 
Table~\ref{tab:kaggle_shadow_aug} shows the overall top1 accuracy for models trained with the KHWD train set. Again, the baseline model is trained without shadow augmentation. The accuracy on the KHWD test set is extremely high for both models with little difference in performance.
However, for testing on the OoD datasets Portable51 and Farm23, we observe improvements for the model trained with shadow augmentation. For both OoD evaluation datasets, we observe around 10\% increase in overall top1 accuracy. However, the degradation in performance is still significant even after applying shadow augmentation. 
Nevertheless, the results still demonstrate the effectiveness of applying shadow augmentation to a dataset without shadow and improving the trained model's performance on unseen datasets with challenging shadow environments. 

\begin{table}[htb!]
\centering
\begin{tabular}{|c|c|c|c|}
\hline
Model            & KHWD-test & Portable51 & Farm23 \\ \hline
Baseline         & 0.961      & 0.281     & 0.264 \\ \hline
Shadow Augmented & 0.964      & 0.308     & 0.291 \\ \hline
Overall Change   & +0.31\%      & \textbf{+9.6\%}     & \textbf{+10.2\%} \\ \hline
\end{tabular}
\caption{Overall accuracy comparison for applying shadow augmentation to training on KHWD dataset.}
\label{tab:kaggle_shadow_aug}
\end{table}

\vspace{-.2cm}
\section{Conclusion}
In this paper, we have explored methods to improve model robustness against varying shadow conditions. First, we have extended our previous investigation on the impact of shadow using system breakdown points. By using a synthetic dataset, we demonstrate that an effective strategy to mitigate the shadow-induced breakdown points is to use additional training data that has heavier and larger shadow. Then we transfer the knowledge and insights gained from synthetic data and develop a shadow augmentation method for training models using real-world data. Through experimental results, we demonstrate the effectiveness of the shadow augmentation method on improving trained model robustness against heavy shadow conditions and distribution shifts across datasets. Also, we compare shadow augmentation with other data augmentation methods and demonstrate superior performance. Our results are verified through multiple trials of using different neural networks and datasets. Future work includes expanding our current shadow augmentation method to incorporate more diversity of shadow and applying it to improve performance on many other tasks that contain outdoor shadow conditions. 

\bibliographystyle{IEEEtran}
\bibliography{ref}

\begin{thebibliography}{10}
\providecommand{\url}[1]{#1}
\csname url@samestyle\endcsname
\providecommand{\newblock}{\relax}
\providecommand{\bibinfo}[2]{#2}
\providecommand{\BIBentrySTDinterwordspacing}{\spaceskip=0pt\relax}
\providecommand{\BIBentryALTinterwordstretchfactor}{4}
\providecommand{\BIBentryALTinterwordspacing}{\spaceskip=\fontdimen2\font plus
\BIBentryALTinterwordstretchfactor\fontdimen3\font minus \fontdimen4\font\relax}
\providecommand{\BIBforeignlanguage}[2]{{%
\expandafter\ifx\csname l@#1\endcsname\relax
\typeout{** WARNING: IEEEtran.bst: No hyphenation pattern has been}%
\typeout{** loaded for the language `#1'. Using the pattern for}%
\typeout{** the default language instead.}%
\else
\language=\csname l@#1\endcsname
\fi
#2}}
\providecommand{\BIBdecl}{\relax}
\BIBdecl

\bibitem{ju2023robust}
S.~Ju, A.~R. Reibman, and A.~J. Deering, ``Robust hand hygiene monitoring for food safety using hand images,'' \emph{Electronic Imaging}, vol.~35, pp. 1--6, 2023.

\bibitem{recht2019imagenet}
B.~Recht, R.~Roelofs, L.~Schmidt, and V.~Shankar, ``Do {I}mage{N}et classifiers generalize to {I}mage{N}et?'' in \emph{International Conference on Machine Learning}.\hskip 1em plus 0.5em minus 0.4em\relax PMLR, 2019, pp. 5389--5400.

\bibitem{ood_bench}
N.~Ye, K.~Li, H.~Bai, R.~Yu, L.~Hong, F.~Zhou, Z.~Li, and J.~Zhu, ``Ood-bench: Quantifying and understanding two dimensions of out-of-distribution generalization,'' in \emph{Proceedings of the IEEE Conference on Computer Vision and Pattern Recognition}, 2022, pp. 7947--7958.

\bibitem{zhong2019hand}
C.~Zhong, A.~R. Reibman, H.~M. Cordoba, and A.~J. Deering, ``Hand-hygiene activity recognition in egocentric video,'' in \emph{2019 IEEE 21st International Workshop on Multimedia Signal Processing}.\hskip 1em plus 0.5em minus 0.4em\relax IEEE, 2019, pp. 1--6.

\bibitem{zhong2020multi}
C.~Zhong, A.~R. Reibman, H.~A. Mina, and A.~J. Deering, ``Multi-view hand-hygiene recognition for food safety,'' \emph{Journal of Imaging}, vol.~6, no.~11, p. 120, 2020.

\bibitem{zhong2021designing}
------, ``Designing a computer-vision application: A case study for hand-hygiene assessment in an open-room environment,'' \emph{Journal of Imaging}, vol.~7, no.~9, p. 170, 2021.

\bibitem{ju2024mipr}
S.~Ju and A.~R. Reibman, ``Exploring the impact of hand pose and shadow on hand-washing action recognition,'' in \emph{IEEE International Conference on Multimedia Information Processing and Retrieval (to appear)}, 2024.

\bibitem{who_food}
``Food safety,'' \url{https://www.who.int/en/news-room/fact-sheets/detail/food-safety}, accessed: 2024-01-28.

\bibitem{liu2019learning}
J.~Liu, H.~Rahmani, N.~Akhtar, and A.~Mian, ``Learning human pose models from synthesized data for robust {RGB-D} action recognition,'' \emph{International Journal of Computer Vision}, vol. 127, pp. 1545--1564, 2019.

\bibitem{lacey_system1}
A.~J. Stewardson, A.~Iten, V.~Camus, A.~Gayet-Ageron, D.~Caulfield, G.~Lacey, and D.~Pittet, ``Efficacy of a new educational tool to improve handrubbing technique amongst healthcare workers: {A} controlled, before-after study,'' \emph{PLoS One}, vol.~9, no.~9, p. e105866, 2014.

\bibitem{lacey_system2}
G.~Lacey, M.~Showstark, and J.~Van~Rhee, ``Training to proficiency in the {WHO} hand hygiene technique,'' \emph{Journal of Medical Education and Curricular Development}, vol.~6, p. 2382120519867681, 2019.

\bibitem{lacey_system3}
G.~Thirkell, J.~Chambers, W.~Gilbart, K.~Thornhill, J.~Arbogast, and G.~Lacey, ``Pilot study of digital tools to support multimodal hand hygiene in a clinical setting,'' \emph{American Journal of Infection Control}, vol.~46, no.~3, pp. 261--265, 2018.

\bibitem{wang2022handwashing}
T.~Wang, J.~Xia, T.~Wu, H.~Ni, E.~Long, J.-P.~O. Li, L.~Zhao, R.~Chen, R.~Wang, Y.~Xu \emph{et~al.}, ``Handwashing quality assessment via deep learning: a modelling study for monitoring compliance and standards in hospitals and communities,'' \emph{Intelligent Medicine}, vol.~2, no.~03, pp. 152--160, 2022.

\bibitem{chinese_journal}
T.~Xie, J.~Tian, and L.~Ma, ``A vision-based hand hygiene monitoring approach using self-attention convolutional neural network,'' \emph{Biomedical Signal Processing and Control}, vol.~76, p. 103651, 2022.

\bibitem{matsushita2002shadow}
Y.~Matsushita, K.~Nishino, K.~Ikeuchi, and M.~Sakauchi, ``Shadow elimination for robust video surveillance,'' in \emph{Workshop on Motion and Video Computing}.\hskip 1em plus 0.5em minus 0.4em\relax IEEE, 2002, pp. 15--21.

\bibitem{inoue2020learning}
N.~Inoue and T.~Yamasaki, ``Learning from synthetic shadows for shadow detection and removal,'' \emph{IEEE Transactions on Circuits and Systems for Video Technology}, vol.~31, no.~11, pp. 4187--4197, 2020.

\bibitem{song2014vehicle}
H.-S. Song, S.-N. Lu, X.~Ma, Y.~Yang, X.-Q. Liu, and P.~Zhang, ``Vehicle behavior analysis using target motion trajectories,'' \emph{IEEE Transactions on Vehicular Technology}, vol.~63, no.~8, pp. 3580--3591, 2014.

\bibitem{shi2020new}
H.~Shi and C.~Liu, ``A new cast shadow detection method for traffic surveillance video analysis using color and statistical modeling,'' \emph{Image and Vision Computing}, vol.~94, p. 103863, 2020.

\bibitem{kaggle}
\BIBentryALTinterwordspacing
A.~Nagaraj, M.~Sood, C.~Sureka, and G.~Srinivasa, ``Sample: Hand wash dataset,'' (accessed September 7, 2021). [Online]. Available: \url{https://www.kaggle.com/datasets/realtimear/hand-wash-dataset}
\BIBentrySTDinterwordspacing

\bibitem{mbnetv3}
A.~Howard, M.~Sandler, G.~Chu, L.-C. Chen, B.~Chen, M.~Tan, W.~Wang, Y.~Zhu, R.~Pang, V.~Vasudevan \emph{et~al.}, ``Searching for {MobileNetV3},'' in \emph{Proceedings of the IEEE International Conference on Computer Vision}, 2019, pp. 1314--1324.

\bibitem{imagenet}
J.~Deng, W.~Dong, R.~Socher, L.-J. Li, K.~Li, and L.~Fei-Fei, ``Imagenet: A large-scale hierarchical image database,'' in \emph{Proceedings of the IEEE Conference on Computer Vision and Pattern Recognition}.\hskip 1em plus 0.5em minus 0.4em\relax Ieee, 2009, pp. 248--255.

\bibitem{resnet}
K.~He, X.~Zhang, S.~Ren, and J.~Sun, ``Deep residual learning for image recognition,'' in \emph{Proceedings of the IEEE Conference on Computer Vision and Pattern Recognition}, 2016, pp. 770--778.

\bibitem{vit}
A.~Dosovitskiy, L.~Beyer, A.~Kolesnikov, D.~Weissenborn, X.~Zhai, T.~Unterthiner, M.~Dehghani, M.~Minderer, G.~Heigold, S.~Gelly \emph{et~al.}, ``An image is worth 16x16 words: Transformers for image recognition at scale,'' \emph{arXiv preprint arXiv:2010.11929}, 2020.

\end{thebibliography}

\end{document}